\documentclass[journal,onecolumn]{IEEEtran}
\usepackage[utf8]{inputenc}
\usepackage[cmex10]{amsmath}
\usepackage{amsfonts,amssymb}
\usepackage{array}
\usepackage{graphicx}
\usepackage[caption=false,font=footnotesize]{subfig}
\usepackage{cite}
\usepackage{algorithm}
\usepackage{epstopdf}
\newcommand{\Int}{\int\limits}

\usepackage[usenames,dvipsnames]{xcolor}

\DeclareMathOperator*{\maxi}{max}

\usepackage{blindtext}
\usepackage{amssymb}
\usepackage{amsthm}

\usepackage{color,soul}
\definecolor{laranja}{rgb}{1.0, 0.6, 0.2}

\newcommand{\at}[2][]{#1|_{#2}}

\begin{document}

\title{On the Secure Energy Efficiency of TAS/MRC with Relaying and Jamming Strategies \\ (Extended Version)}

\author{Jamil~Farhat,~\IEEEmembership{Student~Member,~IEEE,}
		Glauber~Brante,~\IEEEmembership{Member,~IEEE,}
		Richard~Demo~Souza,~\IEEEmembership{Senior~Member,~IEEE}% <-this % stops a space
        \thanks{J. Farhat and G. Brante are with the Federal University of Technology-Paraná, Brazil. E-mails: \{jfarhat@alunos.,gbrante@\}utfpr.edu.br. R. Souza is with the Federal University of Santa Catarina, Brazil, richard.demo@ufsc.br.}%
}

\maketitle

\begin{abstract}
In this paper we investigate the secure energy efficiency (SEE) in a cooperative scenario where all nodes are equipped with multiple antennas. Moreover, we employ secrecy rate and power allocation at Alice and at the relay in order to maximize the SEE, subjected to a constraint in terms of a minimal required secrecy outage probability. Only the channel state information (CSI) with respect to the legitimate nodes is available. Then, we compare the Artificial-Noise (AN) scheme with CSI-Aided Decode-and-Forward (CSI-DF), which exploits the CSI to choose the best communication path (direct or cooperative). Our results show that CSI-DF outperforms AN in terms of SEE in most scenarios, except when Eve is closer to the relay or with the increase of antennas at Eve. Additionally, we also show that the maximization of SEE implies in an optimal number of antennas to be used at each node, which is due to the trade-off between secure throughput and power consumption.
\end{abstract}

\section{Introduction} \label{sec:Introduction}
Physical Layer Security (PLS) schemes have emerged as important tools to enhance confidentiality in wireless systems, adding an extra security layer on top of traditional cryptography by exploiting the fluctuations of the wireless channel~\cite{Bloch}. Among other strategies, multiple antenna (MIMO) schemes have received great interest in the PLS scenario~\cite{Alves.2012, Yang.2013, yang.13b, yang.15}. For instance, the authors in~\cite{Alves.2012} show that Transmit Antenna Selection (TAS) at the legitimate transmitter (Alice) can enhance the secrecy outage probability (SOP) performance with lower cost, complexity and power consumption in comparison with other MIMO strategies, while different combinations of TAS with Maximal Ratio Combining (MRC) or Selection Combining (SC) at the legitimate receiver (Bob) and the eavesdropper (Eve) are investigated in~\cite{Yang.2013, yang.13b, yang.15}. Since only the index of the antenna with the best channel condition is fed back to Alice using TAS, the diversity increases with respect to Bob, but not with respect to Eve, improving security~\cite{Alves.2012}. At Bob, MRC is the optimal strategy in terms of secrecy~\cite{Yang.2013}.

Cooperative protocols have also been intensively investigated in the PLS scenario. Initial studies of the secrecy capacity of cooperative communications have been presented in~\cite{Lai}. Typical cooperative methods employed in the context of PLS are the Decode-and-Forward (DF) and the Artificial-Noise (AN) schemes. The DF protocol combined with TAS at Alice has been analyzed, \emph{e.g.}, in~\cite{Lin.2016} under an outdated channel state information (CSI) feedback assumption showing that the increase of the number of antennas at the legitimate nodes significantly reduces the SOP, even with outdated CSI. Recently, the SOP of the AN scheme has been investigated in~\cite{Benevides.2016}, while a secure energy efficiency (SEE) metric is employed in~\cite{Hu.2016, Jamil.2016}, in which AN is shown to considerably improve the SEE and the SOP when compared to a non-cooperative scheme~\cite{Hu.2016}. Additionally, secure relay and jammer selection using AN against multiple eavesdroppers is studied in~\cite{Hui.2015}. Moreover, the DF, AF and AN schemes were comparatively investigated, considering single antenna devices and a partial security regime in~\cite{Jamil.2016}, where a trade-off between SOP and SEE is established, so that the SEE can be considerably increased by relaxing the security requirement.

In this paper we consider a cooperative MIMO scenario in which Alice employs TAS, due to its good performance, low cost and reduced feedback requirement, while Bob and Eve employ MRC due to its optimality and manageable cost. Moreover, we assume that the relay either employs jamming,  designing a beamforming vector so that the AN affects Eve without interfering at Bob, or cooperates using a CSI-aided DF (CSI-DF) protocol,  in which the legitimate nodes use the available CSI to choose between direct and cooperative paths. In addition, we also employ power and rate allocation aiming at maximizing the SEE, subjected to a maximum SOP constraint. Our results show that CSI-DF outperforms AN in most scenarios, except when Eve is closer to the relay or when the number of antennas at Eve is sufficiently large. Finally, we show that there is an optimum number of antennas at the relay and Bob to maximize the SEE due to the trade-off between secure throughput and power consumption.

% SE SOBRAR ESPAÇO
%Next, Section~\ref{sec:Preliminaries} presents the system model, Section~\ref{sec:MetricasSeguranca} defines the security metrics and Section~\ref{sec:NumericalResults} gives some numerical examples. Finally, Section~\ref{sec:Conclusions} concludes the paper.

\section{System Model}
\label{sec:Preliminaries}
We consider Alice ($\text{A}$), Bob ($\text{B}$) and a relay node ($\text{R}$) communicating in the presence of Eve ($\text{E}$), each one equipped with $n_{\text{A}}$, $n_{\text{B}}$, $n_{\text{R}}$ and $n_{\text{E}}$ antennas, respectively. Assuming TAS at the transmitters $i \in \{\text{A}, \text{R}\}$, the frame received by the $k$-th antenna at any node $j \in \{\text{R}, \text{B}, \text{E}\}$, $i \neq j$, is given by
\begin{equation}
\mathbf{y}_{ij} = \sqrt{\kappa_{ij} P_i} \, h_{ij} \, \mathbf{x} + \mathbf{w}_{j},
\label{ReceivedFrame}
\end{equation}
where $P_i$ is the transmit power of node $i$, $\mathbf{x}$ is the unit average energy transmitted symbol vector, $\mathbf{w}_{j}$ is the zero-mean complex Gaussian noise vector with variance $N_{0}/2$ per dimension and $h_{ij}$ is the quasi-static channel realization, with zero-mean and unit-variance complex Gaussian distributed elements. Moreover, $\kappa_{ij}=\frac{G}{(4 \pi f_\text{c}/c)^2\, d_{ij}^{\upsilon}\, M_\text{l}\, N_\text{f}}$ is the path-loss, where $G$ is the total antenna gain, $f_\text{c}$ is the carrier frequency, $c$ is the speed of light, $d_{ij}$ is the distance between $i$ and $j$, $\upsilon$ is the path-loss exponent, $M_\text{l}$ is the link margin and $N_\text{f}$ is the noise  at the receiver~\cite{Goldsmith}.

In the employed TAS scheme, Bob informs the index of the best transmit antenna through an open feedback channel. The selected antenna at Alice is a random event from the point of view of the relay and Eve, giving no diversity gain. Thus, following~\cite{Yang.2013}, the signal to noise ratio (SNR) at the receiving nodes follows one of two probability density functions (PDFs):  
\begin{equation}
f_{i\text{B}}^{\text{TAS/MRC}}(\gamma) = \frac{n_{i}\,\gamma^{n_{\text{B}}-1}\,e^{-\frac{\gamma}{\bar\gamma}}}{\Gamma(n_{\text{B}})\,{\bar\gamma}^{n_{\text{B}}}}\, 
\left(1-e^{-\frac{\gamma}{\bar\gamma}}\,\sum_{k=0}^{n_{\text{B}}-1} \frac{\gamma^{k}}{k!\bar\gamma^{k}}\right)^{n_{i}-1}
\label{PDF_TAS_MRC}
\end{equation}at Bob, due to the combination of TAS and MRC, and
\begin{equation}
f_{ij}^{\text{MRC}}(\gamma) = \frac{\gamma^{n_{j}-1}\,e^{-\frac{\gamma}{\bar\gamma}}}{\Gamma(n_{j})\,{\bar\gamma}^{n_{j}}}
\label{PDF_MRC}
\end{equation}at the relay and Eve due to MRC only, where $\Gamma(.)$ is the complete gamma function~\cite[\textsection 8.339.1]{gradshteyn.2007}, $i \in \{\text{A}, \text{R}\}$, $j \in \{\text{R}, \text{E}\}$, with $i \neq j$. For the last, we also write its cumulative distribution function (CDF) as~\cite{Lin.2016}
\begin{equation}
F_{ij}^{\text{MRC}}(\gamma) = 1 - e^{-\frac{\gamma}{\bar\gamma}}\,\sum_{w=0}^{n_{j}-1} \left(\frac{1}{w!}\right)\left(\frac{\gamma}{\bar\gamma}\right)^w.
\label{CDF_MRC}
\end{equation}
Moreover, $\bar{\gamma}_{ij} = \frac{\kappa_{ij} P_i}{N_0 B}$ is the average SNR, with $B$ being the system bandwidth, while $\gamma_{ij}$ represents the instantaneous SNR taking the fading realization into account.

\section{Cooperative Schemes} \label{subsection:CooperativeSchemes}
In this section we derive the SOP, defined by~\cite{Bloch} as $p_{\text{out}}^\text{(sch)} = \Pr\left\{C_{\text{B}}-C_{\text{E}}< \mathcal{R}\right\}$, where $\text{sch} \in \{\text{CSI-DF}, \text{AN}\}$, $C_{\text{B}}$ and $C_{\text{E}}$ represent the instantaneous capacities of legitimate and Eve's channel, respectively, and $\mathcal{R}$ is the secrecy rate. 

\subsection{CSI-Aided Decode-and-Forward (CSI-DF)}
Usually, a given channel realization is estimated using a sequence of training symbols sent by the transmitter. Then, the receiver decomposes the obtained channel information $\mathbf{h}$ into the channel direction information (CDI), denoted by $\frac{\mathbf{h}}{||\mathbf{h}||}$, and the channel gain information (CGI), denoted by $||\mathbf{h}||$~\cite{zhang.14, zhang.15}. The CGI is real and positive, so that it can be efficiently quantized with a small number of bits~\cite{yoo.07}, while the CDI is a complex vector with the same number of dimensions as the number of receive antennas~\cite{zhang.14, zhang.15}, thus, quite complex to be quantized. Since we employ TAS at the transmitters, only the CGI is needed, which allows a reduced usage of the feedback, leading to a more practical implementation. Then, with the available CGI, the proposed CSI-DF scheme is also able to choose the most advantageous path: direct or cooperative. If cooperation is chosen, the relay employs the DF scheme to forward the information to Bob at the second time slot. Otherwise, if the direct path is more advantageous, we consider that Alice transmits twice to make the comparison fair. Thus, the capacity of the legitimate channel is given by
\begin{equation}
\begin{split}
C_\text{B}^\text{(CSI-DF)} = &\frac{1}{2} \max\{\log_{2}\left(1+\gamma_\text{dir}\right), \min\left\{\log_{2}\left(1 + \gamma_\text{AR}\right), \log_{2}\left(1 + \gamma_\text{coop}\right)\right\}\},
\end{split}
\label{Capacidade_CSI_RC}
\end{equation}
where $\gamma_\text{coop} = \gamma_\text{AB} + \gamma_\text{RB}$ is the equivalent SNR at Bob when cooperation occurs and $\gamma_\text{dir} = \gamma_\text{AB,1} + \gamma_\text{AB,2}$ is the equivalent SNR when Alice transmits in two consecutive time slots.

At Eve, we consider an optimistic assumption with respect to its channel capacity (thus, pessimistic in terms of secrecy capacity), so that
\begin{equation}
\begin{split}
	C_\text{E}^{(\text{CSI-DF})} = \frac{1}{2} \log_{2}(1+\gamma_\text{AE}+\gamma_\text{RE}),
\end{split}
\label{Capacidade_CSI_EVE_RC}
\end{equation}which assumes that the relay always participates in the legitimate transmission. Let us remark that this assumption is made to allow the computation of a closed form expression to the SOP by using the maximum between i.i.d. random variables.

Then, considering two i.i.d. random variables, $X_{1}$ and $X_{2}$, their maximum and minimum can be written following~\cite{papoulis.2002}
\begin{equation}
	\label{MaxProbability}
	\Pr\left\{\max\left(X_{1}, X_{2} \leq x\right)\right\} = \Pr\left\{X_{1} \leq x\right\}\,\Pr\left\{X_{2} \leq x\right\}
\end{equation}
\begin{equation}
	\begin{split}
	\label{MinProbability}
	\Pr\left\{\min\left(X_{1}, X_{2} \leq x\right)\right\} &= \Pr\left\{X_{1} \leq x\right\} + \Pr\left\{X_{1} > x\right\}\,\Pr\left\{X_{2} \leq x\right\}
	\end{split}
\end{equation}
so that with an approach similar to~\cite{laneman.04}, taking~\eqref{MaxProbability} and~\eqref{MinProbability} into account, and considering that the correlation between $\gamma_\text{dir}$ and $\gamma_\text{coop}$ is treated as in~\cite{laneman.04}, the SOP yields 
\begin{equation} \label{p_out_CSI_RC}
	p_\text{out}^\text{(CSI-DF)} = \mathcal{O}_{\text{ABE}} \left[\mathcal{O}_{\text{ARE}} + \left(1-\mathcal{O}_{\text{ARE}}\right) \mathcal{O}_{\text{BE}}\right],
\end{equation}
where $\mathcal{O}_{\text{ABE}}=\Pr\left\{\log_{2}(1+\gamma_\text{dir})-\log_{2}(1+\gamma_\text{E})<2\mathcal{R}\right\}$, $\mathcal{O}_{\text{ARE}}=\Pr\left\{\log_{2}(1+\gamma_\text{AR})-\log_{2}(1+\gamma_\text{E})<2\mathcal{R}\right\}$ and $\mathcal{O}_{\text{BE}}=\Pr\left\{\log_{2}(1+\gamma_\text{coop})-\log_{2}(1+\gamma_\text{E})<2\mathcal{R}\right\}$.

Next, isolating $\gamma_\text{E}$ for each variable in~\eqref{p_out_CSI_RC} we obtain
\begin{equation}
\begin{split}
p_\text{out}^\text{(CSI-DF)} =& \Pr\{\gamma_\text{E} > 2^{-2\mathcal{R}}\left(1+\gamma_\text{dir}\right)-1\} \Bigg[\Pr\{\gamma_\text{E} > 2^{-2\mathcal{R}}\left(1+\gamma_\text{AR}\right)-1\} \\
&+ \Pr\{\gamma_\text{E} \leq 2^{-2\mathcal{R}}\left(1+\gamma_\text{AR}\right)-1\} 
 \Pr\{\gamma_\text{E} > 2^{-2\mathcal{R}}\left(1+\gamma_\text{coop}\right)-1\} \Bigg],
\end{split}
\label{Integrais_CSI_RC2}
\end{equation}
which requires the PDF related to $\gamma_\text{coop}$ and the CDF related to $\gamma_\text{E}$ to be solved in closed-form.

Considering the convolutions defined by~\cite{Lin.2016}, we can obtain the PDF related to $\gamma_\text{coop}$ and the CDF related to $\gamma_\text{E}$ by doing
\begin{align}
\label{PDF_coop}
f_\text{coop}(x) &= \Int_{0}^{x} f_{\text{RB}}^{\text{TAS/MRC}}\left(x-y\right) f_\text{AB}^{\text{TAS/MRC}}(y)\,\mathrm{d}y, \\
\label{CDF_eve}
F_\text{E}(x) &= \Int_{0}^{x} F_{\text{E,1}}^{\text{MRC}}\left(x-y\right) f_\text{E,2}^{\text{MRC}}(y)\,\mathrm{d}y,
\end{align}
which can be solved with the aid of~\cite[eq. (8)]{Choi.2005},~\cite[\textsection 1.111]{gradshteyn.2007},~\cite[\textsection 3.351.1]{gradshteyn.2007} and~\cite[\textsection 3.381.1]{gradshteyn.2007}, resulting in
\begin{equation}
\begin{split}
	f_\text{coop}(x) &=\frac{n_{\text{R}}n_{\text{A}}\Gamma(n_{\text{B}})^{-2}}{\left(\bar\gamma_{\text{RB}}\bar\gamma_{\text{AB}}\right)^{n_{\text{B}}}} \sum_{k=0}^{n_{\text{A}}-1} \sum_{m=0}^{n_{\text{R}}-1} \binom{n_{\text{A}}-1}{k} \binom{n_{\text{R}}-1}{m} (-1)^{m+k} \prod_{\bar\gamma_{\text{AB}}} \prod_{\bar\gamma_{\text{AR}}}\,e^{-x \left(\frac{k+1}{\bar\gamma_{\text{AB}}}\right)} \\
	&\times x^{n_{\text{B}}+\beta_{1}-1-l}\,\sum_{l=0}^{n_{\text{B}}+\beta_{1}-1}\binom{n_{\text{B}}+\beta_{1}-1}{l}(-1)^{l}\,\left[\left(\frac{m+1}{\bar\gamma_{\text{RB}}}-\frac{k+1}{\bar\gamma_{\text{AB}}}\right)^{-\left(n_{\text{B}}+\beta_{2}+l\right)}\right]\, \\
	&\times \gamma\left(n_{\text{B}}+\beta_{2}+l; \left[\frac{m+1}{\bar\gamma_{\text{RB}}}-\frac{k+1}{\bar\gamma_{\text{AB}}}\right]x\right), \\
\end{split}
\end{equation}
where $\gamma(.)$ is the incomplete gamma function, and
\begin{equation}
\begin{split}
F_\text{E}(x) &= \frac{1}{{\bar\gamma_{\text{AE}}}^{n_{\text{E}}}\Gamma(n_{\text{E}})} \left[\frac{(n_{\text{E}}-1)!}{\bar\gamma_{\text{AE}}^{-n_{\text{E}}}} - e^{-\frac{x}{\bar\gamma_{\text{AE}}}} \sum_{m=0}^{n_{\text{E}}-1} \frac{(n_{\text{E}}-1)!}{m!} \frac{x^m}{\bar\gamma_{\text{AE}}^{m-n_{\text{E}}}} - \sum_{w=0}^{n_{\text{E}}-1} \left(\frac{1}{w!}\right) \left({\frac{1}{\bar\gamma_{\text{RE}}}}\right)^w e^{-\frac{x}{\bar\gamma_{\text{RE}}}} \right. \\
&\left. \sum_{k=0}^{w} \binom{w}{k}\,\left(-1\right)^k\,x^{w-k} \left\{\frac{(k+n_{\text{E}}-1)!}{\left[\frac{1}{\bar\gamma_{\text{AE}}}-\frac{1}{\bar\gamma_{\text{RE}}}\right]^{k+n_{\text{E}}}}-e^{-x \left(\frac{1}{\bar\gamma_{\text{AE}}}-\frac{1}{\bar\gamma_{\text{RE}}}\right)}\,\sum_{m=0}^{k+n_{\text{E}}-1}\frac{\left(k+n_{\text{E}}-1\right)!x^m}{m!\left(\frac{1}{\bar\gamma_{\text{AE}}}-\frac{1}{\bar\gamma_{\text{RE}}}\right)^{k+n_{\text{E}}-m}}\right\}\right].
\end{split}
\end{equation}

With that in hand we can solve~\eqref{p_out_CSI_RC}. Starting with~$\mathcal{O}_{\text{ABE}}$, employing $f_\text{AB}^{\text{TAS/MRC}}$ and $F_{\text{E}}$, we have
\begin{align}
\mathcal{O}_{\text{ABE}}=\Int_{0}^{\infty} F_\text{E}\left[2^{-2\mathcal{R}}\left(1+\gamma_\text{dir}\right)-1\right]f_{\text{AB}}^{\text{TAS/MRC}}\left(\gamma_\text{dir}\right) \,\mathrm{d}{\gamma_\text{dir}}
\end{align}
which, after using the binomial expansion and applying~\cite[\textsection 3.35.3]{gradshteyn.2007}, \cite[eq. (8)]{Choi.2005} and some algebraic manipulations, results in
\begin{align}	
	\mathcal{O}_{\text{ABE}}&=\frac{n_{\text{A}}}{\Gamma(n_{\text{B}})\:\Gamma(n_{\text{E}})\ {\bar\gamma_{\text{AB}}}^{n_{\text{B}}}\ \bar\gamma_{\text{AE}}^{n_{\text{E}}}}\ \sum_{k=0}^{n_{\text{A}}-1} \binom{n_{\text{A}}-1}{k}\ (-1)^{k}\; \prod_{\bar\gamma_{\text{AB}}} \nonumber
	\times \left\{\frac{(n_{\text{E}}-1)!}{\bar\gamma_{\text{AE}}^{-n_{\text{E}}}} \mathcal{X}(0, 0, \frac{k+1}{\bar\gamma_{\text{AB}}}) \right. \\ 
	&- \left. e^{-\frac{2^{-2\mathcal{R}}-1}{\bar\gamma_{\text{AE}}}} \sum_{m=0}^{n_{\text{E}}-1} \frac{m!\Gamma(n_{\text{E}})}{\bar\gamma_{\text{AE}}^{m-n_{\text{E}}}} \times  \sum_{v=0}^{m} \binom{m}{v} \mathcal{X}\left(m, v, \phi(\bar\gamma_{\text{AE}})\right) - \sum_{w=0}^{n_{\text{E}}-1} \sum_{v=0}^{w} \frac{(-1)^{v}}{w!} \left(\frac{1}{\bar\gamma_{\text{RE}}}\right)^w \right. \nonumber\\
	&\left. \times  \left[\:\:e^{-\frac{2^{-2\mathcal{R}}-1}{\bar\gamma_{\text{RE}}}}\:\:\:\sum_{z=0}^{w-v}\:\:\: \binom{w}{v}\ \mathcal{Z}(0)\:\:\mathcal{X}(w-v, z, \phi(\bar\gamma_{\text{RE}}))\:\:\:   - \:\: e^{-\frac{2^{-2\mathcal{R}}-1}{\bar\gamma_{\text{AE}}}}\:\:\sum_{z=0}^{w-v} \sum_{m=0}^{v+n_{\text{E}}-1} \:\:
	\right.\right. \nonumber \\
	&\left.\left. \:\:\sum_{y=0}^{m}\:\:\binom{w}{v}\:\:\binom{m}{y}\:\:\mathcal{Z}(m) \mathcal{X}(w+m-v, z+y, \phi(\bar\gamma_{\text{AE}})) \right. \bigg] \right\},
\label{Probability_CSI_RC_AB}
\end{align}
where $
\phi(a)=\frac{k+1}{\bar\gamma_{\text{AB}}}+\frac{2^{-2\mathcal{R}+1}}{a}$,
$
\mathcal{Z}(x)=\frac{(v+n_{\text{E}}-1)!}{x!\left(\frac{\bar\gamma_{\text{RE}}-\bar\gamma_{\text{AE}}}{\bar\gamma_{\text{RE}}\bar\gamma_{\text{AE}}}\right)^{v+n_{\text{E}}-x}}$, $\mathcal{X}(a, b, c)=\frac{(2^{-2\mathcal{R}}-1)^{a-b}(b+\beta_{1}+n_{\text{B}}-1)!}{2^{b(-2\mathcal{R}+1)}c^{(b+\beta_{1}+n_{\text{B}})}}$ and
\begin{equation}
\label{ProductGammaAB}
\prod_{\bar\gamma_{\text{AB}}}=\prod_{i=1}^{n_{\text{B}}-1} \left[\sum_{u_{i}=0}^{u_{i-1}} \binom{u_{i-1}}{u_{i}}\left({\frac{1}{i!}}\right)^{u_{i}-u_{i+1}}\left(\frac{1}{{\bar\gamma_{\text{AB}}}}\right)^{u_{i}}\right]
\end{equation}
with $\beta_{1}=\sum_{i=1}^{n_{\text{B}}-1} u_{i}$, $u_{0}=k$ and $u_{n_{\text{B}}}=0$.

Next, solving $\Pr\{\gamma_\text{E} > 2^{-2\mathcal{R}}\left(1+\gamma_\text{AR}\right)-1\}$ yields an integral form as
\begin{align}
\mathcal{O}_{\text{ARE}}=\Int_{0}^{\infty} F_\text{E}\left[2^{-2\mathcal{R}}\left(1+\gamma_\text{AR}\right)-1\right]\,f_{\text{AR}}^{\text{MRC}}\left(\gamma_{\text{AR}}\right) \,\mathrm{d}\gamma_{\text{AR}},
\label{Probability_CSI_RC_AR2}
\end{align}
which can be solved in closed-form by substituting $f_{\text{AR}}^{\text{MRC}}$ and $F_\text{E}$, while using an algebraic approach similar to that in~\eqref{Probability_CSI_RC_AB}. After these steps we arrive at
\begin{align}
	\mathcal{O}_{\text{ARE}}&=\frac{\Gamma(n_{\text{E}})^{-1}\Gamma(n_{\text{R}})^{-1}}{\bar\gamma_{\text{AE}}^{n_{\text{E}}}\bar\gamma_{\text{AR}}^{n_{\text{R}}}} \left[\sum_{m=0}^{n_{\text{E}}-1}\frac{\left(n_{\text{E}}-1\right)!}{m! e^{\frac{2^{-2\mathcal{R}}-1}{\bar\gamma_{\text{AE}}}}} \left(\frac{1}{\bar\gamma_{\text{AE}}}\right)^{m-n_{\text{E}}}\: \mathcal{J}(m, \bar\gamma_{\text{AE}})+ \sum_{w=0}^{n_{\text{E}}-1} \left(\frac{1}{w!}\right) \left(\frac{1}{\bar\gamma_{\text{RE}}}\right)^w  \right. \nonumber \\
	&\left. \sum_{k=0}^{w} \binom{w}{k} \left(-1\right)^{k}\,\Big\{\mathcal{T}(\bar\gamma_{\text{RE}}, 0)\:\: \mathcal{J}(w-k, \bar\gamma_{\text{RE}}) - \sum_{o=0}^{k+n_{\text{E}}-1} \mathcal{T}(\bar\gamma_{\text{AE}}, o)\:\: \times \mathcal{J}(w-k+o, \bar\gamma_{\text{AE}}) \Big\}\right].
\label{Probability_CSI_RC_AR}
\end{align}
where
$\mathcal{T}(a, b)=\frac{b!\left(k+n_{\text{E}}-1\right)! e^{-\frac{2^{-2\mathcal{R}}-1}{a}}}{\left[\frac{1}{\bar\gamma_{\text{AE}}}-\frac{1}{\bar\gamma_\text{RE}}\right]^{k+n_{\text{E}}-b}}$ and $
\mathcal{J}(a, b)=\sum_{p=0}^{a}\binom{a}{p} \frac{\left(2^{-2\mathcal{R}}-1\right)^{a-p}\left(n_{\text{R}}+p-1\right)!}{2^{2\mathcal{R}p}}\left(\frac{2^{-2\mathcal{R}}\bar\gamma_{\text{AR}}+b}{\bar\gamma_{\text{AR}}b}\right)^{-\left(n_{\text{R}}+p\right)}$.

Finally, $\mathcal{O}_{\text{BE}}$ can be solved with the aid of $ F_\text{E}$ and $f_{\text{coop}}$, so that
\begin{align}
\mathcal{O}_{\text{BE}}=&\Int_{0}^{\infty} F_\text{E}\left[2^{-2\mathcal{R}}\left(1+\gamma_{\text{coop}}\right)-1\right]\,f_{\text{coop}}\left(\gamma_{\text{coop}}\right) \,\mathrm{d}\gamma_{\text{coop}},
\end{align}
whose closed-form expression is obtained using~\cite[\textsection 6.455.2]{gradshteyn.2007} and some algebraic manipulations, resulting in
\begin{align}	
	\mathcal{O}_{\text{BE}}&=\frac{n_{\text{R}}n_{\text{A}}\left(\bar\gamma_{\text{AB}}\bar\gamma_{\text{RB}}\right)^{-n_{\text{B}}}}{\bar\gamma_{\text{AE}}^{n_{\text{E}}}\Gamma(n_{\text{E}})\Gamma(n_{\text{B}})^{2}} \sum_{k=0}^{n_{\text{A}}-1} \sum_{m=0}^{n_{\text{R}}-1} \binom{n_{\text{A}}-1}{k} \binom{n_{\text{R}}-1}{m} \prod_{\bar\gamma_{\text{AB}}} \prod_{\bar\gamma_{\text{AR}}} \sum_{l=0}^{n_{\text{B}}-\beta_{1}-1} \frac{\binom{n_{\text{B}}+\beta_{1}-1}{l}\left(-1\right)^{k+m+l}}{\left(\frac{m+1}{\bar\gamma_{\text{RB}}}-\frac{k+1}{\bar\gamma_{\text{AB}}}\right)^{\left(n_{\text{B}}+\beta_{2}+l\right)}} \nonumber\\
	& \times  \left[\frac{\Gamma(n_{\text{E}})}{\bar\gamma_\text{AE}^{-n_{\text{E}}}}  \mathcal{B}\left(0, \mu(0), \psi(0)\right) -\:\:\sum_{p=0}^{n_{\text{E}}-1} \frac{\Gamma(n_{\text{E}})\:\:e^{-\frac{2^{-2\mathcal{R}}-1}{\bar\gamma_\text{AE}}}\:\:\bar\gamma_\text{AE}^{n_{\text{E}}-p}}{p!}\:\mathcal{B}\left(p, \mu(s), \psi\left(\frac{2^{-2\mathcal{R}}}{\bar\gamma_{\text{AE}}}\right)\right) \right. \nonumber\\
	&\left. - \sum_{w=0}^{n_{\text{E}}-1} \left(\frac{1}{w!}\right) \left(\frac{1}{\bar\gamma_{\text{RE}}}\right)^w \sum_{o=0}^{w} \binom{w}{o} \left(-1\right)^o \times \bigg\{ \mathcal{C}(\bar\gamma_{\text{RE}}, 0) \mathcal{B}\left(w-o, \mu(s), \psi\left(\frac{2^{-2\mathcal{R}}}{\bar\gamma_{\text{RE}}}\right)\right) \right. \nonumber\\
	&\left. + \mathcal{C}(\bar\gamma_{\text{AE}}, z) \mathcal{B}\left(w+z-o, \mu(s), \psi\left(\frac{2^{-2\mathcal{R}}}{\bar\gamma_{\text{AE}}}\right)\right) \bigg\}\right],
\label{Probability_CSI_RC_B}
\end{align}
where 
$\mathcal{B}(a, b, c)= \sum_{s=0}^{a} \binom{a}{s} \frac{\left(2^{-2\mathcal{R}}-1\right)^{a-s}}{2^{2\mathcal{R}s}} \frac{\alpha^v \Gamma(b+v)}{v\left(\alpha+c\right)} {}_2F_1\left(1;b+v;v+1;\frac{\alpha}{\alpha+c}\right)$,
with $_{2}F_{1}\left(\alpha,\beta;\gamma;z\right)$ being the Gauss hypergeometric function~\cite[\textsection 9.111]{gradshteyn.2007}, and $\alpha=\frac{\bar\gamma_{\text{AB}}(m-1)-\bar\gamma_{\text{RB}}(k+1)}{\bar\gamma_{\text{AB}}\bar\gamma_{\text{RB}}}$, $v=n_{\text{B}}+\beta_{2}+l$, $\mu(a)=a+n_{\text{B}}+\beta_{1}-l$, $\psi(a)=\frac{k+1}{\bar\gamma_{\text{AB}}}+a$,
$\mathcal{C}(a, b)=\sum_{i=0}^{b} \frac{\left(o+n_{\text{E}}-1\right)!e^{-\frac{2^{-2\mathcal{R}}-1}{a}}}{b!\left[\frac{\bar\gamma_\text{RE}-\bar\gamma_\text{AE}}{\bar\gamma_\text{AE}\bar\gamma_\text{RE}}\right]^{o+n_{\text{E}}-b}}$,
\begin{align}
\prod_{\bar\gamma_{\text{AR}}}=\prod_{i=1}^{n_{\text{R}}-1} \left[\sum_{v_{i}=0}^{v_{i-1}} \binom{v_{i-1}}{v_{i}}\left({\frac{1}{i!}}\right)^{v_{i}-v_{i+1}}\left(\frac{1}{{\bar\gamma_{\text{AR}}}}\right)^{v_{i}}\right], 
\label{ProductGammaAR}
\end{align}
with $\displaystyle \beta_{2}=\sum_{i=1}^{n_{\text{R}}-1} v_{i}$, $v_{0}=m$ and $v_{n_{\text{R}}}=0$.

Then, the closed-form expression to the SOP of CSI-DF is obtained after applying~\eqref{Probability_CSI_RC_AB},~\eqref{Probability_CSI_RC_AR} and~\eqref{Probability_CSI_RC_B} into~\eqref{p_out_CSI_RC}.

\subsection{Artificial-Noise (AN)}
%Similar to~\cite{Benevides.2016}, we consider a scenario where Alice employs TAS and Bob employs MRC while the communication to Eve is degraded by multiple interfering signals. However, differently from~\cite{Benevides.2016}, we consider that these interfering signals are generated by the multiple antennas of the relay, which creates a beamforming vector so that the noise is null in the direction of Bob. {\color{blue} An important assumption is that the number of antennas at the relay must be larger than the number of antennas at Bob~\cite{Benevides.2016, Zhu.2013}; thus, $n_\text{B} \leq n_\text{R}-1$ is always considered for the AN scheme.} Then, the SOP for such scheme yields~\cite{Benevides.2016}
Several works in the literature have shown that multiple antennas can increase the PLS. A notable strategy is to use jamming in order to confuse Eve, as \emph{e.g.} in~\cite{Benevides.2016}, in which Alice employs TAS and Bob employs MRC while the communication to Eve is degraded by multiple interfering signals. However, differently from~\cite{Benevides.2016}, we consider that these interfering signals are generated by the multiple antennas of the relay, which creates a beamforming vector so that the noise is null in the direction of Bob. Thus, the jamming affects only Eve, without interfering at Bob. An important assumption with respect to the creation of the beamforming vector is that the number of antennas at the relay must be larger than the number of antennas at Bob~\cite{Benevides.2016, Zhu.2013}; thus, $n_\text{B} \leq n_\text{R}-1$ is always considered for the AN scheme. 

Then, the capacity of the legitimate channel is given by
\begin{equation} 
C_\text{B}^\text{(AN)} = \frac{1}{2} \log_{2}\left(1+\gamma_\text{AB}\right),
\label{Capacidade_AN}
\end{equation}
while at Eve the capacity is limited by the jamming (denoted by~\cite{Benevides.2016} as interference) generated by the relay node. Thus, we represent the signal-to-interference ratio (SIR) at Eve following the same notation of~\cite{Benevides.2016}, so that $\Upsilon_{\text{I}}=\frac{\gamma_\text{AE}}{\gamma_\text{I}}$, where $\displaystyle \gamma_\text{I} = \sum_{k=0}^{n_{\text{R}}} \bar\gamma_{\text{RE}, k}\,|h_{\text{RE}, k}|^2$ is the jamming interference, written as the sum of all jamming signals sent by each $k$-th antenna of the relay. As a result, the capacity of Eve's channel can be written as
\begin{equation} \label{Capacidade_Eve_AN}
	C_\text{E}^{(\text{AN})} = \frac{1}{2} \log_{2}\left(1+\Upsilon_{\text{I}}\right).
\end{equation}

Therefore, the secrecy outage probability for the AN scheme can be written as
\begin{equation} \label{p_out_AN}
	\begin{split}
	p_\text{out}^\text{(AN)} &= \Pr\left\{\frac{1+\gamma_\text{AB}}{1+\Upsilon_{\text{I}}} < 2^{2\mathcal{R}}\right\}= \Int_{0}^{\infty} F_\text{AB}^{\text{TAS/MRC}}\left[2^{2\mathcal{R}}\left(1+x\right)-1\right]f_{\frac{\gamma_\text{AE}}{\gamma_\text{I}}}\left(x\right) \,\mathrm{d}{x},
	\end{split}
\end{equation}
where $\displaystyle F_\text{AB}^{\text{TAS/MRC}}(z)=\left[1 - e^{-\frac{\gamma_{\text{AB}}}{\bar\gamma_{\text{AB}}}}\,\sum_{w=0}^{n_{\text{B}}-1} \left(\frac{1}{w!}\right)\left(\frac{\gamma_{\text{AB}}}{\bar\gamma_{\text{AB}}}\right)^w\right]^{n_{\text{A}}}$ is given by~\cite[eq. (13)]{Benevides.2016} and,
\begin{equation} \label{PDF_GammaI}
	\begin{split}
	f_{\frac{\gamma_\text{AE}}{\gamma_\text{I}}}\left(x\right)={\frac{\partial}{\partial x}}\left[\Int_{0}^{\infty} F_{\text{AE}}^{\text{MRC}}\left(xz\right) f_{\gamma_\text{I}}\left(z\right)\,\mathrm{d}{z} \right], 
	\end{split}
\end{equation}
with $\displaystyle F_{\text{AE}}^{\text{MRC}}(\gamma_\text{AE}) = 1 - e^{-\frac{\gamma_\text{AE}}{\bar\gamma_\text{AE}}}\,\sum_{w=0}^{n_{\text{E}}-1} \left(\frac{1}{w!}\right)\left(\frac{\gamma_\text{AE}}{\bar\gamma_\text{AE}}\right)^w$ and $\displaystyle f_{\gamma_\text{I}}\left(z\right)=\sum_{i=1}^{n_{\text{R}}}\,\frac{e^{-\frac{z}{\bar\gamma_{\text{I}}}}}{\bar\gamma_{\text{I}}}$, following~\cite[eq. (19)]{Benevides.2016}. In addition, let us remark that we consider the power is equally distributed among the jamming signals to simplify the analysis.

Then, the SOP for the AN scheme yields~\cite{Benevides.2016}
\begin{align}
p_\text{out}^\text{(AN)}=&1- \sum_{k=1}^{n_{\text{A}}} \binom{k}{n_{\text{A}}} \prod_{\bar\gamma_{\text{AB}}} \sum_{u=0}^{n_{\text{E}}-1} \frac{(-1)^{k+1}}{u!} \frac{\Gamma(u+n_{\text{R}})}{\Gamma(n_{\text{R}})} \sum_{p=0}^{\beta_{1}} \binom{\beta_{1}}{p} \left(\frac{\bar\gamma_{\text{AE}}}{\bar\gamma_{\text{RE}}}\right)^{p} \left(2^{2\,\mathcal{R}}-1\right)^{\beta_{1}-p} 2^{2\,\mathcal{R}p} e^{-\frac{k(z-1)}{\bar\gamma_{\text{AB}}}} \left[ n_{\text{R}} \times \right. \nonumber \\
	 &\left. \times \Gamma(p+u+1) \Psi\left(p+u+1, p-n_{\text{R}}+1, \frac{k\bar\gamma_\text{AE}2^{2\,\mathcal{R}}}{\bar\gamma_\text{RE}\bar\gamma_\text{AB}}\right)-\mathcal{L}\right],
	 \label{SOPAN}
\end{align}
where 
\begin{align}
\mathcal{L}=\begin{cases} 
u\Gamma(p+u)\Psi\left(p+u, p-n_{\text{R}}, \frac{k\bar\gamma_{\text{AE}}2^{2\,\mathcal{R}}}{\bar\gamma_\text{RE}\bar\gamma_\text{AB}}\right)\, & \mbox{if } u \neq 0
\\ 0, & \mbox{if } u=0 \end{cases} \nonumber
\end{align}
with $\Psi(.,.,.)$ denoting the Tricomi's confluent hypergeometric function~\cite[\textsection 9.211.4]{gradshteyn.2007}. Let us remark that~\eqref{SOPAN} is associated with $n_{\text{B}}$ due to the term $\prod_{\bar\gamma_{\text{AB}}}$ defined in~\eqref{ProductGammaAB}.

\subsection{Secure Energy Efficiency and Optimization} \label{sec:MetricasSeguranca}
In order to capture both security and energy efficiency issues, let us define the SEE metric~as
\begin{equation}
\eta_\text{s} = \frac{\mathcal R\, \left(1-p_{\text{out}}^\text{(sch)}\right)}{P_{\text{total}}^\text{(sch)}},
\label{EnergyEfficiency}
\end{equation}
where $P_\text{total}^\text{(sch)}$ is the total power consumed by each cooperative scheme. In the case of CSI-DF we have
\begin{align}
&P_{\text{total}}^\text{(\text{CSI-DF})} = 2 \left[(1+\delta)P_{\text{A}} + P_{\text{TX}} + n_{\text{B}}\,P_{\text{RX}}\right]\, p_\text{dir} \\
&+\left[(1+\delta)(P_{\text{A}}+P_{\text{R}}) + 2\,P_{\text{TX}} + \left(2\,n_{\text{B}}+n_{\text{R}}\right) P_{\text{RX}} \right] p_\text{coop}, \nonumber
\label{TotalPowerCSI_RC}
\end{align}
where $P_{\text{TX}}$ and $P_\text{RX}$ denote the power consumed by transmission and reception circuitry, respectively, while the efficiency loss of the power amplifier is $\delta$. Moreover, $p_\text{coop}$ ($p_\text{dir}$) is the probability that the transmission is cooperative (direct), given by $p_\text{coop} = 1 - p_\text{dir} \approx \frac{\bar\gamma_\text{AR}}{\bar\gamma_\text{AR} + \bar\gamma_\text{AB}}$~\cite{Jamil.2016}.

On the other hand, in the case of AN we have
\begin{equation}
P_{\text{total}}^\text{(AN)} = 2\left[(1+\delta)\,(P_{\text{A}} + P_{\text{R}}) + \,\left(n_{\text{R}}+1\right)\,P_{\text{TX}} + n_\text{B}\,P_{\text{RX}}\right].
\label{TotalPowerAN}
\end{equation}

Additionally, we are interested in maximizing the SEE of each scheme by allocating $P_\text{A}$ and $P_\text{R}$, as well as the secrecy rate. The general optimization problem can be formalized as
\begin{equation}
\begin{split}
\maxi_{(P_\text{A}, P_\text{R}, \mathcal{R})} \qquad & \eta_\text{s}^{(\text{sch})} = \frac{\mathcal R\, \left(1-p_{\text{out}}^\text{(sch)}\right)}{P_{\text{total}}^{(\text{sch})}} \\
\text{s.t.} \qquad  & 0 < P_{i} \leq P_{\text{max}},\; \text{with} \; i \in \{\text{A}, \text{R}\}, \\
& 0 \leq \mathcal{R} \leq \mathcal{R}_{\text{max}}, \\
& p_{\text{out}}^{(\text{sch})} \leq \varphi,
\label{OptimizationEE}
\end{split}
\end{equation}
where $P_{\text{max}}$ is a maximum transmit power constraint, $\varphi$ is the maximum acceptable SOP of the system and $\mathcal{R}_{\text{max}} = C_{\text{B}}-\mathcal{R}_{\text{E}}$ is the maximum secrecy rate, assuming that $ C_{\text{B}}$ is known and that the target equivocation rate is given by $\mathcal{R}_{\text{E}}$.

An alternative to solve~\eqref{OptimizationEE} is related to the use of low-complex iterative algorithms such as the Dinkelbach algorithm~\cite{Zappone.2015, Jamil.2016} to perform the power allocation, while the optimal secrecy rate can be obtained, \emph{e.g.}, by the golden section search algorithm. Let us remark that a similar approach has been used by the authors in the past~\cite{Jamil.2016}, so that we elaborate a brief explanation of the proposed algorithms in the following.

The Dinkelbach algorithm~\cite{dinkelbach.67, Zappone.2015} allows to optimize the ratio between functions of the same variable (fractional programming). Therefore, the algorithm is specially useful for the optimization of the powers allocated at Alice and at the relay. A fractional programming is represented in a general form by~\cite{dinkelbach.67}
\begin{equation}
	\maxi_{\mathbf{x \in S}} \; q(x)=\frac{f_{1}(x)}{f_{2}(x)},
\label{NonLinearFractional}
\end{equation}
where $\mathbf{S}\subseteq \mathbb{R}^{n}$ is a convex set, $f_{1}$, $f_{2}:\mathbf{S} \to \mathbb{R}$, being $f_{1}(x)$ concave and $f_{2}(x)>0$ convex. Using a parametric convex program, it is possible rewrite~\eqref{NonLinearFractional} as~\cite{Zappone.2015}
\begin{equation}
F(\lambda)=\maxi_{\mathbf{x \in S}} \; f_{1}(x)-\lambda f_{2}(x),
\label{FractionalProgramI}
\end{equation}
in which $f_{1}(x)$ is maximized while $f_{2}(x)$ is minimized, with the parameter $\lambda$ being the weight associated with the denominator. Moreover, the optimum value of the function is found with
\begin{equation}
F(\lambda)=0 \iff  \lambda = q^\star,
\label{RelacaoOtima}
\end{equation}
where $q^\star$ is the optimum value of~\eqref{NonLinearFractional}. Therefore, solving~\eqref{NonLinearFractional} is equivalent to finding the root of
\begin{equation}
F(\lambda^\star)=\maxi_{\mathbf{x \in S}} f_{1}(x)-\lambda f_{2}(x) = 0.
\label{OptimalCondition}
\end{equation}
	
The Dinkelbach algorithm is an efficient approach to find the solution of~\eqref{OptimalCondition}, which is based on Newton's method to find $\lambda$ for each $(n+1)$-th iteration by doing
\begin{equation}
\lambda_{n+1} = \lambda_{n}-\frac{F(\lambda_{n})}{F'(\lambda_{n})} = \frac{f_{1}(x_{n}^\star)}{f_{2}(x_{n}^\star)}.
\label{MetodoNewton}
\end{equation}

Due to the allocation of the power at Alice and at the relay, the power allocation can be splitted in two steps: starting with Alice and then allocating power to the relay. Therefore, with respect to Alice, \eqref{OptimalCondition} can be rewritten as
\begin{equation}
F_{1}(\lambda)=\maxi_{P_{\text{A}} \geq 0} f_{1}(P_{\text{A}})-\lambda f_{2}(P_{\text{A}}) = 0,
\label{F1_Dink}
\end{equation}
where $f_{1}(P_{\text{A}})=\mathcal R\, \left(1-p_{\text{out}}^\text{(sch)}\right)$ and $f_{2}(P_{\text{A}})=P_{\text{total}}^\text{(sch)}$. Moreover, the stationary condition is given by
\begin{equation}
{\frac{\partial f_{1}(P_{\text{A}})}{\partial P_{\text{A}}}}\at[\Bigg]{P_{\text{A}}=P_{\text{A}}^\star} - \lambda \,\, {\frac{\partial f_{2}(P_{\text{A}})}{\partial P_{\text{A}}}}\at[\Bigg]{P_{\text{A}}=P_{\text{A}}^\star} = 0,
\label{DerivativeF1}
\end{equation}
with $P_{\text{A}}^\star$ obtained by the Dinkelbach method~\footnote{As we can observe, the complexity of the method is inherent to the complexity of the SOP expressions for each cooperative scheme, due to the derivatives specified in~\eqref{DerivativeF1} and~\eqref{DerivativeF2}. However, although the SOP expressions are complex for arbitrary number of antennas at each node, the expressions can be considerably simplified while fixing $n_{\text{A}}$, $n_{\text{B}}$, $n_{\text{R}}$ and $n_{\text{E}}$.}.

In the sequence, considering the power allocation to the relay we have the
\begin{equation}
F_{2}(\lambda)=\maxi_{P_{\text{R}} \geq 0} f_{1}(P_{\text{R}})-\lambda f_{2}(P_{\text{R}}) = 0,
\label{F2_Dink}
\end{equation}
where $f_{1}(P_{\text{R}})=\mathcal R\, \left(1-p_{\text{out}}^\text{(sch)}\right)$, $f_{2}(P_{\text{R}})=P_{\text{total}}^\text{(sch)}$ and with the following stationary condition
\begin{equation}
{\frac{\partial f_{1}(P_{\text{R}})}{\partial P_{\text{R}}}}\at[\Bigg]{P_{\text{R}}=P_{\text{R}}^\star} - \lambda \,\, {\frac{\partial f_{2}(P_{\text{R}})}{\partial P_{\text{R}}}}\at[\Bigg]{P_{\text{R}}=P_{\text{R}}^\star} = 0,
\label{DerivativeF2}
\end{equation}
with $P_{\text{R}}^\star$ also obtained by the Dinkelbach method.

Finally, with respect to the optimization of $\mathcal{R}$, we can employ a golden section search algorithm with parabolic interpolation as in~\cite{Jamil.2016}. Such algorithm allows to find the maximum of an unimodal function by narrowing the range of values inside a predefined interval~\cite{Press2007_NumericalRecipes}.

\section{Numerical Results}
\label{sec:NumericalResults}
We consider that $\mathcal{R}=3$~bps/Hz, $d_\mathrm{AB} = 100$~m, $\upsilon=3$ and $\varphi=10^{-1}$. Moreover, as in~\cite{Jamil.2016}, $P_{\mathrm{TX}} = 112.2$~mW, $P_{\mathrm{RX}} = 97.9$~mW, $B=10$~kHz, $N_{0} = -174$~dBm/Hz, $M_\text{l} = 40$~dB, $G = 5$~dBi, $N_\text{f} = 10$~dB and $f_\text{c}=2.5$~GHz.

\begin{figure}[!t] % normalmente utilizar [!t]
	\centering
	\includegraphics[width=10cm]{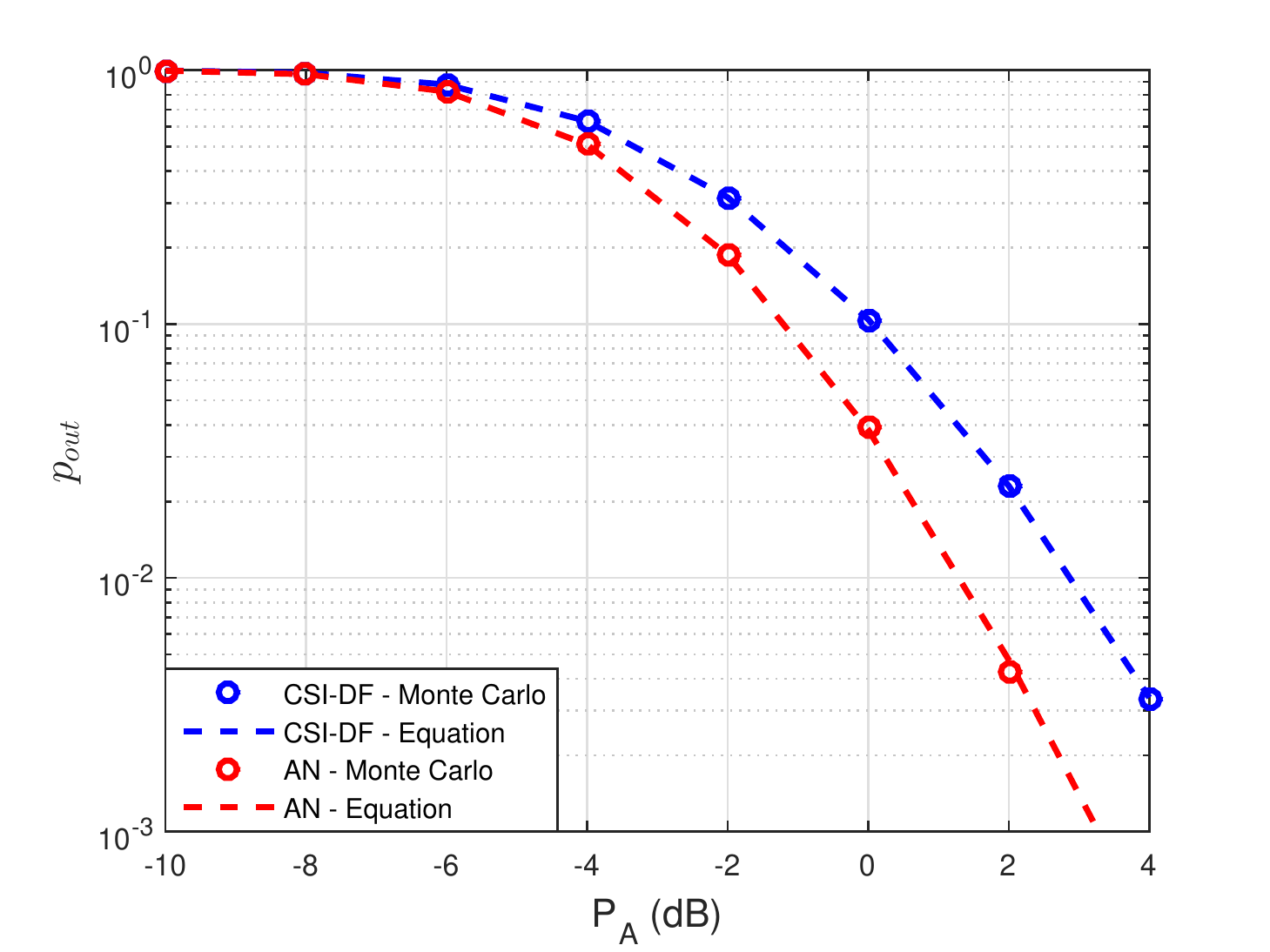}
	\caption{Secrecy outage probability of CSI-DF and AN schemes, comparing exact expressions and Monte Carlo simulations for $d_\mathrm{AR} = 0.5\,d_\mathrm{AB}$.}
	\label{Outage}
\end{figure}

First, Fig.~\ref{Outage} plots the secrecy outage probability of CSI-DF and AN, comparing both the exact expressions and Monte Carlo simulations. Although the figure only considers the case when the relay is placed an a intermediate position between Alice and Bob ($d_\text{AR}=0.5\,d_\text{AB}$), the same agreement between theoretic and simulation results is observed for different positions of the relay.

\begin{figure}[!t] % normalmente utilizar [!t]
\centering
\includegraphics[width=10cm]{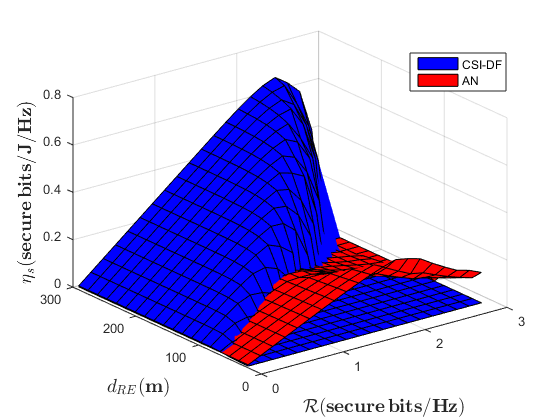}
\caption{SEE of CSI-DF and AN as a function of $\mathcal{R}$ and $d_\text{RE}$, with $d_\text{AR}=0.5\,d_\text{AB}$ and $n_\text{A} = n_\text{R} = n_\text{E} = 2$ with $n_\text{B}=1$ for AN scheme.}
\label{EnergyEfficiency_SecrecyRate_dRE}
\end{figure}

Fig.~\ref{EnergyEfficiency_SecrecyRate_dRE} plots the SEE as a function of $\mathcal{R}$ and $d_{\text{RE}}$, where we observe that, when Eve is closer, AN performs better since the relay interferes with more intensity at Eve, increasing the SEE. On the other hand, CSI-DF allows important improvements when $\mathcal{R}$ increases, also leading to the highest SEE point. 

\begin{figure}[!t] % normalmente utilizar [!t]
	\centering
	\includegraphics[width=10cm]{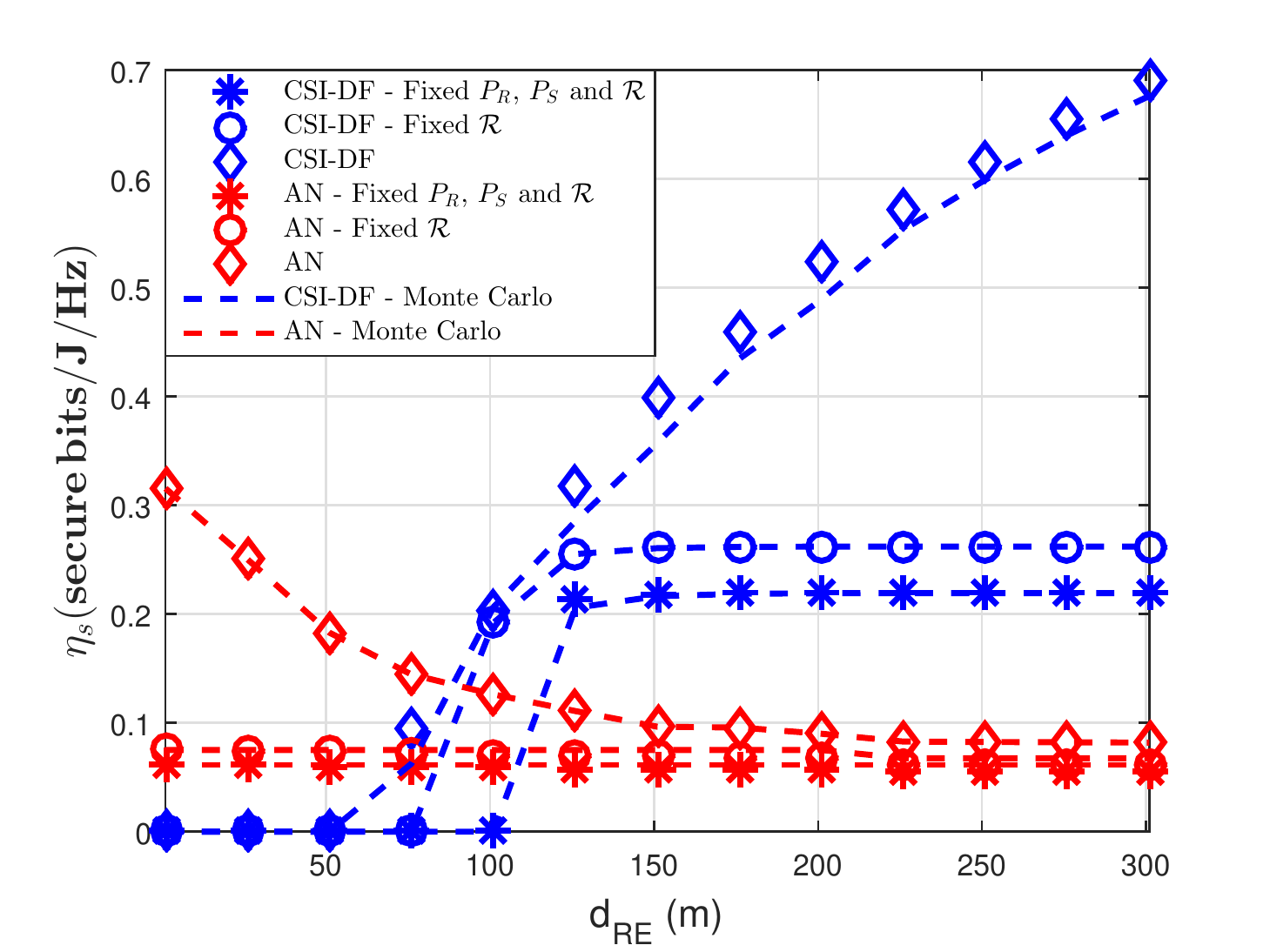}
	\caption{SEE of CSI-DF and AN as a function of $d_\text{RE}$ for different allocation strategies, with $d_\text{AR}=0.5\,d_\text{AB}$ and $n_\text{A} = n_\text{R} = n_\text{E} = 2$ with $n_\text{B}=1$.}
	\label{EnergyEfficiency_dRE}
\end{figure}

Next, Fig.~\ref{EnergyEfficiency_dRE} compares the SEE with fixed power and rate, fixed rate and power allocation, and the rate and power allocation defined by~\eqref{OptimizationEE}. Additionally, we compare the SEE expressions with Monte Carlo simulations, in which a good agreement is shown. As we observe, a significant performance improvement is obtained when power and rate are allocated. In particular, it is worth noting that power allocation plays a major role to maximize the SEE. 

\begin{figure}[!t] % normalmente utilizar [!t]
\centering
\includegraphics[width=10cm]{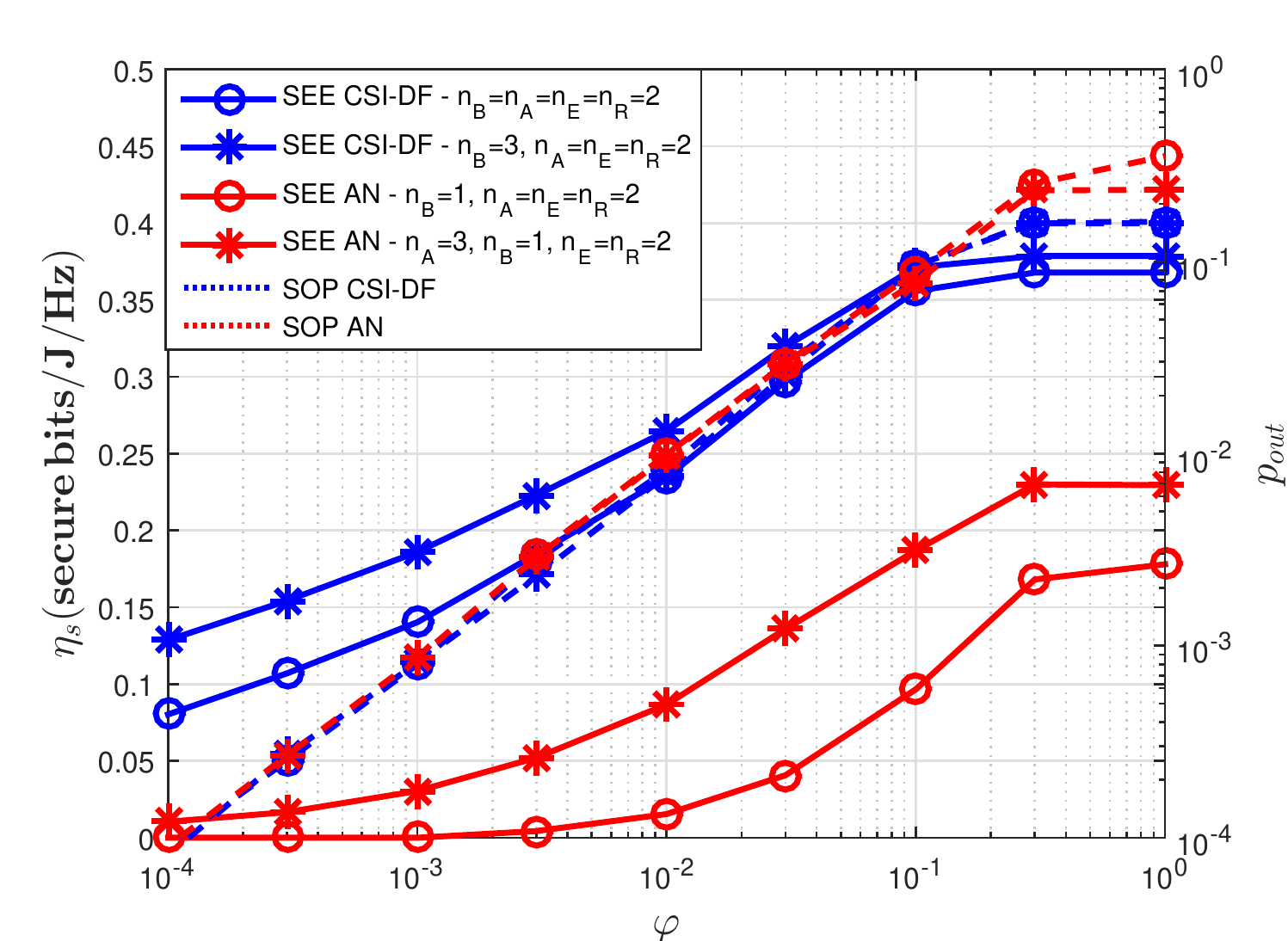}
\caption{SEE, represented by solid lines, and SOP, represented by dashed lines, of CSI-DF and AN as a function of $\varphi$, with $d_\text{RE}=1.5\,d_\text{AB}$, $d_\text{AR}=0.5\,d_\text{AB}$.}
\label{EnergyEfficiency_OutageAlvo}
\end{figure}

Fig.~\ref{EnergyEfficiency_OutageAlvo} plots the SEE, represented by solid lines, and the SOP, represented by dashed lines, as a function of $\varphi$, the minimal requirement for the SOP, which shows that a higher $\varphi$ increases the SEE, despite the penalty related to the number of secure transmitted bits. However, let us remark that even with an increase of SEE with $\varphi$, the SOP that maximizes the SEE is not close to one. This occurs due to the fact that a SOP close to one implies in a SEE that tends to zero in~\eqref{EnergyEfficiency}.

\begin{figure}[!t] % normalmente utilizar [!t]
	\centering
	\includegraphics[width=10cm]{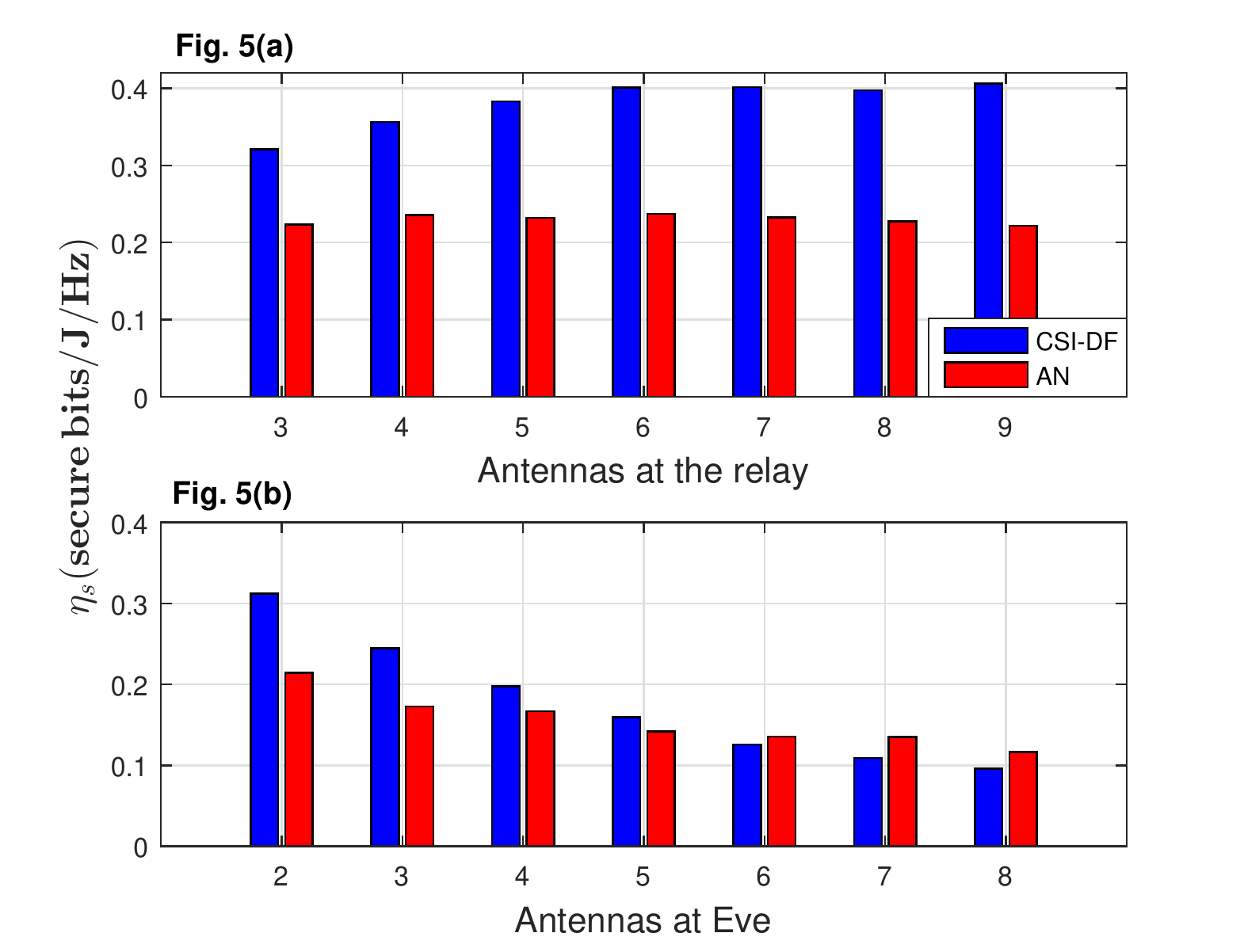}
	\caption{SEE of CSI-DF and  AN as a function of the number of antennas at the relay and at Eve for $d_\text{AR}=0.5\,d_\text{AB}$, $d_\text{RE}=1.25\,d_\text{AB}$.}
	\label{EnergyEfficiency_Antennas}
\end{figure}

Finally, Fig.~\ref{EnergyEfficiency_Antennas} illustrates the SEE as a function of the number of antennas, with $n_\text{A} = n_{\text{B}}=n_{\text{E}}=2$ while varying $n_\text{R}$ in Fig.~\ref{EnergyEfficiency_Antennas}(a), and while varying $n_\text{E}$ with fixed $n_{\text{B}}=n_{\text{A}}=2$ and $n_{\text{R}}=3$ in Fig.~\ref{EnergyEfficiency_Antennas}(b). As we observe, increasing $n_\text{R}$ is more advantageous to CSI-DF than to AN, once the increase of $n_\text{R}$ yields a diversity gain in the case of CSI-DF when cooperation occurs. On the other hand, increasing $n_\text{R}$ in the AN scheme only yields a larger power consumption. In addition, in Fig.~\ref{EnergyEfficiency_Antennas}(b) we can observe that AN becomes more advantageous with the increase of $n_\text{E}$, \emph{i.e.}; when the number of antennas at Eve is much larger than that of the legitimate nodes, it is better for the relay to interfere at Eve by injecting Gaussian noise rather than to cooperate with Bob.

\section{Conclusions}
\label{sec:Conclusions}
We investigate the SEE in a cooperative MIMO scenario with different setups with respect to the number of antennas and the maximum acceptable SOP of the system, also considering power and secrecy rate allocation. We compare a CSI-DF scheme, which exploits the available CSI to chose between direct or cooperative transmission, with an AN scheme, in which the relay uses a beamforming vector to interfere only at Eve. Results show that CSI-DF outperforms AN in most scenarios, except if Eve is closer to the relay or with the increase of antennas at Eve, when AN becomes more advantageous.

\bibliographystyle{IEEEtran}
\bibliography{IEEEabrv,article}
\end{document}